\documentclass{prop2015}
\usepackage[english]{babel}
\usepackage{t1enc}
\usepackage[utf8]{inputenc}
\usepackage[english]{babel}
\normalfont
\usepackage{amsmath}
\usepackage{amssymb}
\usepackage{bbm}
\usepackage{bm}
\usepackage{hyperref}
\usepackage{pgf}
\usepackage{graphicx}
\usepackage[inline]{enumitem}

\newcommand{\be}[0]{\begin{equation}}
\newcommand{\ee}[0]{\end{equation}}

\title{A modern point of view on anomalies}
\author[S. Monnier]{Samuel Monnier\inst{1,}\footnote{Corresponding author\quad E-mail:~\textsf{samuel.monnier@gmail.com}}}
\address[1]{Section de Math\'ematiques, Universit\'e de Gen\`eve, 2-4 rue du Li\`evre, 1211 Gen\`eve 4, Switzerland}
\shortauthors{S. Monnier}
\begin{abstract}
We review the concept of anomaly field theory, namely the fact that the anomalies of a $d$-dimensional field theory can be encoded in a $d+1$-dimensional field theory functor. We give numerous examples of anomaly field theories, explain how classical facts about anomalies are recovered from the anomaly field theory, and review recent work on global anomaly cancellation in 6d supergravity where this concept was instrumental. We also sketch the status of global anomaly cancellation checks in string theory. This paper is based on a talk given at the Durham Symposium "Higher Structures in M-theory" in August 2018. 
\end{abstract}
\shortabstract
\begin{document}
\maketitle

\section{Introduction}

\label{Intro}

The aim of this review paper is to explain the concept of \emph{anomaly field theory}, which offers a convenient way of packaging all the anomalies of a quantum field theory into a single well-defined mathematical object.

In the rest of the present section, we explain why consistency constraints are so important in string theory, as well as how anomalies yield such consistency constraints. We review the Atiyah-Segal picture of quantum field theories as functors in Section \ref{FieldThFunc}, and the concept of relative field theory in Section \ref{AnomRelFieldTh}. We explain that relative field theories are essentially anomalous field theories, taking values in their respective anomaly field theories. We give a few examples of pairs of anomalous/anomaly field theories in Section \ref{Examples}. In Section \ref{ClassPictAnom}, we show how anomaly field theories implement the well-known properties of anomalies. Section \ref{UseCase} provides a concrete example where the concept of anomaly field theory plays a crucial role. In Section \ref{AnomCanStringTh}, we conclude with a review of solved and open problem pertaining to global anomaly cancellation in string theory backgrounds.

Most of the material in the present paper is exposed in more detail in \cite{Monnierd}.

\paragraph{The unreasonable power of self-consistency} Practical constraints restrain the type of experiments we can perform to test our understanding of the world. In particular, barring unlikely scenarios, quantum gravity effects are expected to be relevant at energies close to the Planck scale, well out of reach of current or foreseeable experiments. A legitimate question is then: Why do string theorists have a hope of constructing a fundamental theory of quantum gravity, given that there is direct experimental input on this problem? This question is asked repeatedly by outsiders, but seems to rarely get an appropriate answer, with negative consequences for the image of string theory among the general public.

String theorists believe they can construct a fundamental theory of quantum gravity because self-consistency overconstrains string theory. This not only leads to a unique way of constructing the theory, but provides in addition powerful consistency tests. Epistemologically, these tests are no different from experimental tests, in the sense that their irremediable failure would rule out string theory as a theory of quantum gravity. This very fact is what makes string theory hard science rather than a bundle of conjectures.

For a historical perspective, it is useful to remember how general relativity was constructed. Requiring that the laws of mechanics and electromagnetism hold in accelerated reference frames, Einstein was able to derive general relativity purely from consistency considerations. He was lucky that general relativistic effects happened to be within experimental reach, allowing the theory to be tested experimentally right away. We do not seem to have such luck with string theory, but the methodology consisting in constructing the theory from self-consistency is completely analogous. Of course, this kind of situation is rather exceptional in physics. The typical situation is illustrated by particle physics: the realm of quantum field theories contains an infinite number of consistent theories, and experimental data was absolutely necessary to select the Standard Model over other consistent quantum field theories.

In this context, an important task is to understand in great detail the consistency constraints that string theory should satisfy. There are many sources of such constraints, but we will focus here on anomaly cancellation.

\paragraph{Anomalies as consistency conditions} Suppose that a classical theory $\mathcal{C}$ admits $G$ as a symmetry. It may happen that the quantization process cannot be carried out without explicitly breaking $G$, and that the $G$ symmetry fails to be restored in the resulting quantum theory $\mathcal{Q}$. Invariant observables of $\mathcal{C}$ only have covariant expectations values in $\mathcal{Q}$, in the sense that the latter transform in non-trivial representations of $G$\footnote{Indeed, the anomaly is due to a lack of invariance of the integration measure used to compute expectations \cite{Fujikawa:1979ay}.}. We then say that the symmetry $G$ has an \emph{anomaly}, or is \emph{anomalous}, in $\mathcal{Q}$. Similarly, we say that $\mathcal{Q}$ has an \emph{anomaly}, or is \emph{anomalous} if the symmetry of interest is anomalous in $\mathcal{Q}$.

A fundamental construction in physics is \emph{gauging}, which can be thought of as the physically sensible way of quotienting a theory by the action of a symmetry $G$. It involves coupling the theory to a background gauge field in order to make a global symmetry local, restricting to the invariant observables, identifying gauge equivalent configurations of the gauge field and promoting it to a dynamical field. But in an anomalous theory, the invariant observables generally have non-invariant expectation values, which is obviously incompatible with the identification of gauge equivalent configurations. This means that in the presence of an anomaly, while the classical gauged theory exists, it has no quantum counterpart; anomalies are obstructions to the existence of quantum gauge theories. There are associated consistency constraints on quantum field theories and string theory: any gauged symmetry should be non-anomalous.

\paragraph{Anomalies from geometric invariants} It is well-known that the local anomalies of a $d$-dimensional quantum field theory are encoded in a degree $d+2$-dimensional characteristic form, which for chiral fermions is the degree $d+2$ index density of a certain Dirac operator. The characteristic form does not however contain information about global anomalies, i.e. anomalies of symmetries not connected to the identity. Global (as well as local) anomalies can be recovered from a certain $d+1$-dimensional geometric invariant. The invariant is "geometric" because it can depend on geometric data, such as a Riemannian metric or a background connection. In the case of chiral fermions, this geometric invariant is the exponential of the modified eta invariant (see Section \ref{Examples} below for details).

\begin{figure}
  \includegraphics[width=\columnwidth]{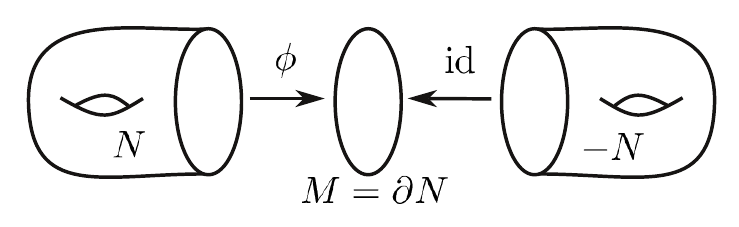}%
  \caption{\label{FigGlobAnComp} The prescription described in the main text to compute anomalous phases.
   }
\end{figure}

Given the geometric invariant $I$, the anomalous phase picked by the partition function of the theory under a symmetry transformation $\phi$ is computed as follows (see Figure \ref{FigGlobAnComp}). Let $M$ be the spacetime and let $N$ be a $d+1$-dimensional manifold with boundary admitting $M$ as a component of $\partial N$. (Depending on the situation, it may not be possible to find a $d+1$-dimensional manifold having $M$ as its only boundary component, but a $N$ as above always exists.) Let $-N$ be $N$ with its orientation flipped. Glue $N$ to $-N$ along their respective boundary using $\phi$ to identify the $M$ boundary components of $N$ and $-N$, and the identity on the other boundary components, to obtain a closed $d+1$-dimensional manifold $\tilde{N}$. The sought-for anomalous phase is then $I(\tilde{N})$. The reason for this prescription and its compatibility with the local anomaly will become clear in Section \ref{ClassPictAnom}.

\section{Field theories as functors}

\label{FieldThFunc}

\paragraph{Setup and conventions} We will always consider the Euclidean, Wick-rotated versions of the quantum field theories of interest, and restrict ourselves to compact spacetime manifolds. This not a problem when studying anomalies, because any lack of invariance should survive the analytic continuation relating the Euclidean and Lorentzian theories. For simplicity, we will always assume that the spacetimes on which the quantum field theory of interest lives is oriented.

Quantum field theories typically require topological and geometric structures on the spacetimes on which they are defined. We will call these structures the \emph{field theory data} of the quantum field theory under consideration. For instance, the theory of a free spin $1/2$ field on oriented spacetime requires a spin stucture and a Riemannian metric. More generally, the field theory data can be abstracted as a sheaf over spacetime. In the following, "manifold" will always mean "manifold endowed with the relevant field theory data". 

Given a manifold, we will often find it convenient to write the dimension and the highest corner codimension as an exponent: $M^{d,k}$ is a $d$-dimensional manifold with codimension $k$ corners. In particular $M^{d,1}$ is a $d$-dimensional manifold with boundary, and $M^d := M^{d,0}$ is a closed $d$-dimensional manifold.

\begin{figure}
  \includegraphics[width=\columnwidth]{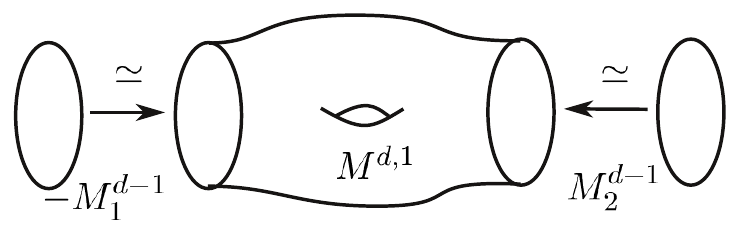}%
  \caption{\label{FigBordism} A bordism $M^{d,1}$ from $M^{d-1}_1$ to $M^{d-1}_2$.
   }
\end{figure}

\paragraph{Bordism categories} A $d$-dimensional bordism is a manifold $M^{d,1}$ with boundary endowed with a partition of its boundary as $\partial M^{d,1} \simeq -M^{d-1}_1 \sqcup M^{d-1}_2$, where $-$ denotes the orientation flip and $\sqcup$ is the disjoint union (Figure \ref{FigBordism}). Given bordisms, we can construct a bordism category $\mathcal{B}^d$ (up to a subtlety to be discussed soon) by seeing $d-1$-dimensional manifolds as objects and $d$-dimensional bordisms $M^{d,1}$, $\partial M^{d,1} \simeq -M^{d-1}_1 \sqcup M^{d-1}_2$, as morphisms from $M^{d-1}_1$ to $M^{d-1}_2$. Composition is given by gluing bordisms along common boundary components.

In order to really obtain a category from the construction above, we need the objects in $\mathcal{B}^d$ to be endowed with a $d$-dimensional germ of the sheaf encoding field theory data on spacetime, see Appendix A.4 of \cite{Monnierd} for details. The gluing operations should be compatible with the germs involved.

\begin{figure}
  \includegraphics[width=\columnwidth]{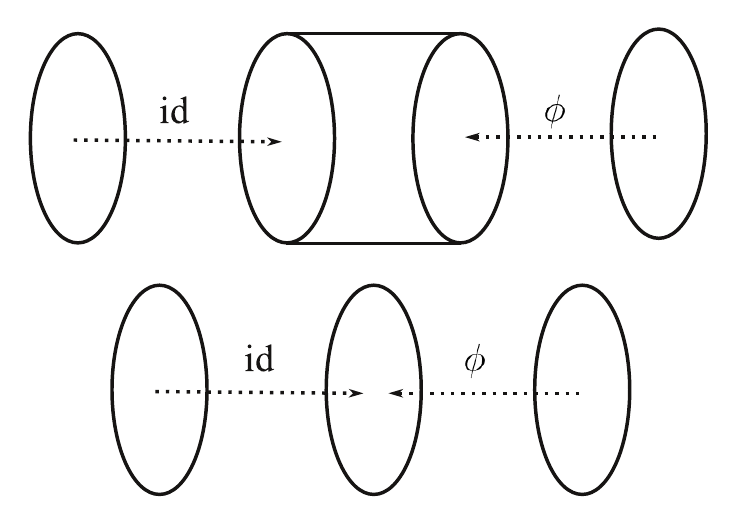}%
  \caption{\label{FigInfBord} Top: A "twisted cylinder" $C = M \times S^1$, with the left boundary identified with $M$ through the identity map and the right boundary identified with $M$ through an automorphism $\phi$. The twisted cylinder is a bordism from $M$ to $\phi M$. In the limit where the length of the cylinder tends to zero, we end up with an infinitesimal bordism. The infinitesimal bordisms generate an action of the automorphism group of $M$ on any bordism with a boundary component isomorphic to $M$.
   }
\end{figure}

Moreover, there may not be any bordism corresponding to the identity map. This occurs in particular when the field theory data includes a Riemannian metric: gluing any bordism to another strictly increases the $d$-dimensional volume, showing that the identity map cannot be represented in this way. The solution is to include as well formal "infinitesimal" bordisms $\bar{\phi}$ for each isomorphism $\phi$ of the field theory data on $M^{d-1}$. $\bar{\phi}$ is then a morphism from $M^{d-1}$ to $\phi(M^{d-1})$, given concretely by the isomorphism $\phi$. See Figure \ref{FigInfBord} for a geometrical interpretation of the infinitesimal bordism $\bar{\phi}$. The infinitesimal bordism associated to the identity automorphism ${\rm id}_{M^{d-1}}$ is then the identity morphism at $M^{d-1}$ in $\mathcal{B}^d$. The other infinitesimal bordisms are also very useful to describe the symmetries of quantum field theories, as we will see. More details about bordism categories can be found in Appendix A.4 of \cite{Monnierd}.

Bordism categories carry a natural symmetric monoidal structure, given by the disjoint union of manifolds.

\paragraph{Atiyah-Segal axioms} The Atiyah-Segal axioms are a convenient way of formalizing the notion of quantum field theory into a mathematical object. In this framework, a $d$-dimensional quantum field theory $\mathcal{Q}$ is a symmetric monoidal functor from a bordism category $\mathcal{B}^d$ to the category of Hilbert spaces $\mathcal{H}$. $\mathcal{B}^d$ should be the bordism category of manifolds endowed with the field theory data required by $\mathcal{Q}$. 

The symmetric monoidal property requires that the empty $d-1$-dimensional manifold $\emptyset^{d-1}$ is always associated to the 1-dimensional Hilbert space $\mathbb{C}$, and the empty $d$-dimensional bordism $\emptyset^d$ is associated to the identity homomorphism $\mathbb{C} \stackrel{1}{\rightarrow} \mathbb{C}$. This implies that a closed $d$-dimensional manifold $M^d$ (a bordism $\emptyset^{d-1} \rightarrow \emptyset^{d-1}$) is associated to a homomorphism $\mathbb{C} \rightarrow \mathbb{C}$, which corresponds canonically to a complex number. $\mathcal{Q}(M^d)$ is interpreted as the Euclidean partition function of $\mathcal{Q}$ on $M^d$. $\mathcal{Q}(M^{d-1})$ is a Hilbert space in $\mathcal{H}$, and is interpreted as the state space of $\mathcal{Q}$ on $M^{d-1}$.

\paragraph{The trivial theory} Given a bordism category $\mathcal{B}^d$, a very simple functor $\bm{1}^d$ into $\mathcal{H}$ can be constructed by assigning to every object the vector space $\mathbb{C}$ and to any morphism the identity $\mathbb{C} \stackrel{1}{\rightarrow} \mathbb{C}$. The corresponding field theory, the \emph{trivial $d$-dimensional field theory}, is too simple to be physically interesting, but will nevertheless play a central role in the following.

\paragraph{Invertible theories} A quantum field theory $\mathcal{Q}$ is said to be \emph{invertible} if the corresponding functor takes "invertible" values. For instance, the state space $\mathcal{Q}(M^{d-1})$ should be an invertible Hilbert space with respect to the monoidal structure of $\mathcal{H}$ given by the tensor product, i.e. a 1-dimensional Hilbert space. It is also required that $\mathcal{Q}(M^d) \in \mathbb{C}^\ast$, and more generally that $\mathcal{Q}(M^{d,1})$ (which is a homomorphism between two 1-dimensional Hilbert spaces) should be invertible.

The trivial $d$-dimensional theory is an example of an invertible quantum field theory. 

\paragraph{Symmetries} Suppose that an isomorphism $\phi: M^{d-1} \rightarrow \phi M^{d-1}$ is a symmetry of $\mathcal{Q}$, meaning that the state spaces $\mathcal{Q}(M^{d-1})$ and $\mathcal{Q}(\phi M^{d-1})$ are canonically isomorphic. We use this canonical isomorphism to identify them. $\phi$ corresponds to a limit bordism $\bar{\phi}: M^{d-1} \rightarrow \phi M^{d-1}$ in $\mathcal{B}^d$. A quantum field theory $\mathcal{Q}$, being a functor from $\mathcal{B}^d$ to $\mathcal{H}$, will associate a morphism $Q(\bar{\phi}): \mathcal{Q}(M^{d-1}) \rightarrow \mathcal{Q}(\phi M^{d-1}) = \mathcal{Q}(M^{d-1})$ to $\bar{\phi}$. $Q(\bar{\phi})$ describes the action of the symmetry $\phi$ on the state space of $\mathcal{Q}$ on $M^{d-1}$. The axioms defining functors also ensure that given a symmetry group $G$ of $\mathcal{Q}$ on $M^{d-1}$, the automorphisms $\{\mathcal{Q}(\bar{\phi})\}_{\phi \in G}$ form a representation of $G$ on $\mathcal{Q}(M^{d-1})$. 

\paragraph{Extended field theories} Physical quantum field theories display interesting structures and observables associated to submanifolds of codimension larger than 1, generally known as defects. These structures are not captured by the model above in term of functors from a bordism category to the category of Hilbert spaces. They can be included by generalizing $\mathcal{Q}$ to a functor between higher categories. Indeed, there is a bordism $k$-category consisting of manifolds of dimension $d-k$ to $d$ with corners of dimension at least $d-k$. Similarly, there is a $k$-category of $k$-Hilbert spaces. An \emph{extended field theory} to codimension $k$ is a symmetric monoidal $k$-functor from a $k$-category of bordism into the $k$-category of Hilbert spaces.

In order to capture all the observables of a quantum field theory, we should set $k = d$, i.e. consider fully extended field theory functors. This is often not practical because of the difficulty of representing concretely higher categories. We will nevertheless have to consider extended field theories down to codimension $2$. A bordism 2-category has as objects closed $d-2$-dimensional manifolds, as 1-morphisms $d-1$-dimensional manifolds with boundary and as 2-morphisms $d$-dimensional manifolds with corners. An ordinary bordism category can be recovered as the category of morphisms over the empty $d-2$-dimensional manifold $\emptyset^{d-2}$. The 1-morphisms from $\emptyset^{d-2}$ to itself, which are closed $d-1$-dimensional manifolds, are interpreted as the objects of the ordinary bordism category, while the 2-morphisms between such morphisms, which are $d$-dimensional manifolds with boundary, are interpreted as the morphisms of the ordinary bordism category.

Extended field theory functors associate a 2-Hilbert space to closed $d-2$-dimensional manifolds. A finite 2-Hilbert space of dimension $n$ is a category linearly equivalent to $\mathcal{H}^n$, the $n$th Cartesian product of the category of (finite) Hilbert spaces with itself, endowed with some extra structure categorifying the inner product of Hilbert spaces (see Appendix A.2 of \cite{Monnierd}). Through such an equivalence, the objects of a 2-Hilbert space can be pictured as "vectors of Hilbert spaces". To our knowledge, the notion of infinite-dimensional 2-Hilbert space has not been developed yet. Fortunately, we will see that using finite-dimensional 2-Hilbert spaces only still allows us to discuss many quantum field theories of physical interest. The role of $\mathbb{C}$ in the category of Hilbert spaces is now played by the category $\mathcal{H}$ itself, seen as a 1-dimensional 2-vector space. In particular, the trivial 2-extended $d$-dimensional field theory assigns $\mathcal{H}$ to any $d-2$-dimensional closed manifold.

\section{Anomalies and relative field theories}

\label{AnomRelFieldTh}

\paragraph{Relative field theories} In the Atiyah-Segal picture, quantum field theories are symmetric monoidal functors from a bordism category $\mathcal{B}^d$ into the category of Hilbert spaces $\mathcal{H}$. A natural question is: what do natural transformations between such functors correspond to?

Let us consider a natural transformation $\eta$ between two copies of the trivial field theory functor $\bm{1}^d$. $\eta$ can be seen as an assignment of a morphism $\eta_{M^{d-1}}: \bm{1}^d(M^{d-1}) \rightarrow \bm{1}^d(M^{d-1})$ in $\mathcal{H}$ to each object $M^{d-1}$ of $\mathcal{B}^{d-1}$. But $\bm{1}^d(M^{d-1}) = \mathbb{C}$, so $\eta_{M^{d-1}}: \mathbb{C} \rightarrow \mathbb{C}$ can be canonically pictured as a complex number. 

One can repeat the analysis by considering $\bm{1}^d$ as an extended field theory down to codimension 2. A 2-natural transformation then essentially amounts to a family of functors $\eta_{M^{d-2}}: \bm{1}^d(M^{d-2}) \rightarrow \bm{1}^d(M^{d-2})$ indexed by closed $d-2$-dimensional manifolds (the objects of the 2-category $\mathcal{B}^{d-2}$). But $\bm{1}^d(M^{d-2}) = \mathcal{H}$, and it turns out that a functor $\eta_{M^{d-2}}: \mathcal{H} \rightarrow \mathcal{H}$ can always be represented as the tensor product operation with a fixed Hilbert space $H_{M^{d-2}}$. 

We see that a (2-)natural transformation $\eta: \bm{1}^d \rightarrow \bm{1}^d$ assigns a Hilbert space to each $d-2$-dimensional closed manifold $M^{d-2}$ and a complex number to each closed $d-1$-dimensional manifold $M^{d-1}$. This suggests that $\eta$ is equivalent to a $d-1$-dimensional field theory functor. For this to be true, we actually have to restrict the functors $\bm{1}^d$ to the bordism category of $d-2$-dimensional manifolds with $d-1$-dimensional bordisms, an operation that we write $|_{d-1}$. One can then show natural transformations between two copies of the restricted $d$-dimensional trivial field theory $\bm{1}^d|_{d-1}$ are $d-1$-dimensional field theory functors, see Section 3.1 of \cite{Monnierd} for the full argument. In the following, in order to avoid cluttering in the notation, we will omit the restriction operation, but it will always be understood.

This new way of picturing field theory functors is interesting because it admits an obvious generalization: instead of considering natural transformations between copies of the trivial field theory functor, we can consider natural transformations between any pair of (suitably restricted) field theory functors. It turns out that nothing is gained by having two general field theory functors and that one can always choose, say, the target of the natural transformation to be the trivial field theory functor.  

A $d$-dimensional \emph{relative quantum field theory}  \cite{Freed:2012bs} is therefore defined as a (2-)natural transformation $\mathcal{R}: \mathcal{A} \rightarrow \bm{1}^{d+1}$, where $\mathcal{A}: \mathcal{B}^{d+1} \rightarrow \mathcal{H}$ is a $d+1$-dimensional field theory functor.

Intuitively, relative field theories are best pictured as "field theories taking value in a field theory $\mathcal{A}$". By working out again the definition of a natural transformation, one finds that $\mathcal{R}(M^d)$ is simply a vector in $\mathcal{A}(M^d)$. So the "partition function" of $\mathcal{R}$ on $M^d$ is actually a vector in the state space of $\mathcal{A}$ on $M^d$. Similarly, the "state space" of $\mathcal{R}$ on $M^{d-1}$ is an object in the 2-Hilbert space $\mathcal{A}(M^{d-1})$ assigned by $\mathcal{A}$ to $M^{d-1}$.

\paragraph{Anomalous = relative} In \cite{Freed:2014iua}, Freed suggested that $d$-dimensional anomalous field theories should be pictured as relative field theories $\mathcal{F}: \mathcal{A} \rightarrow \bm{1}^{d+1}$, where $\mathcal{A}$ is a $d+1$-dimensional field theory, the \emph{anomaly field theory}. (Similar ideas appeared in unpublished work by Moore \cite{Moore, Moore2012}, as well as in the condensed matter literature \cite{Wen:2013oza, Kong:2014qka}.) The partition function of the anomaly field theory on closed $d+1$-dimensional manifolds coincides with the geometric invariant $I$ mentioned in Section \ref{Intro}.

We will justify extensively this proposal in Section \ref{ClassPictAnom}, by recovering well-known properties of anomalous field theories from the relative field theory framework. Let us emphasize that the original proposal in \cite{Freed:2014iua} assumed $\mathcal{A}$ to be an \emph{invertible} field theory functor. While the traditional examples of anomalous field theories do indeed fall into this category, we do not see a good reason to impose invertibility. As the examples will show, the invertible and non-invertible cases have many similarities, and a form of anomaly cancellation is also possible in the non-invertible case.

\section{Examples}

\label{Examples}

\paragraph{Complex chiral fermions}

Recall that in dimension $d = 4k + 2$, there are two non-isomorphic irreducible spinor modules in Lorentzian signature. As a consequence massless chiral fermions exist in these dimensions. It has been known for a long time that chiral fermions are anomalous \cite{PhysRev.177.2426, Fujikawa:1979ay}. 

Chiral fermionic fields are (anti-commuting) sections of the bundle $S_\pm \otimes V$, where $S_\pm$ is the positive/negative chirality spinor bundle over spacetime and $V$ is a \emph{twist} bundle determined by the field theory data, in which the fermionic fields are valued. The geometric invariant computing global anomalies is the following. On a $4k+3$-dimensional manifold, the field theory data determines a corresponding twist bundle $V$ and we can consider the Dirac operator $D_V$ twisted by $V$. The eta invariant $\eta_V$ of $D_V$ is a suitably regularized version of the number of positive eigenvalue minus the number of negative eigenvalues (which are both infinite). The modified eta invariant is defined by $\xi_V := \frac{1}{2} (\eta_V + h)$, where $h$ is the (finite) number of zero eigenvalues of $D_V$. The geometric invariant computing the anomalies of the chiral fermions valued in $V$ is $\exp 2\pi i \xi_V$ \cite{Witten:1985xe, MR861886} (see Section \ref{Intro}).

According to the discussion of the preceding section, the anomaly field theory of chiral fermions should be a field theory functor whose partition function on closed $4k+3$-dimensional manifolds is the exponentiated modified eta invariant $\exp 2\pi i \xi_V$. Exactly such a field theory functor was constructed by Dai and Freed in \cite{Dai:1994kq} (see also Section 9 of \cite{Freed:2016rqq} for a construction using homotopy theory and Section 5 of \cite{Monnierd} for the extension to codimension 2).  The state space of this theory on a closed $d$-dimensional manifold is the determinant line of the chiral Dirac operator $D$ defining the chiral fermionic theory (i.e. the fiber of its determinant line bundle \cite{0603.32016}). This means that the partition function of the chiral fermions is an element of this determinant line. This is completely consistent with the well-known fact that that, over the space of background fields, the partition function of chiral fermions is not a complex function, but rather a section of the determinant line bundle of $D$, see for instance \cite{Freed:1986zx}. 

\paragraph{Majorana fermions} Majorana and symplectic Majorana fermions are defined from a complex fermionic theory by requiring invariance under a certain involution, and contain half the corresponding degrees of freedom. Keeping the same notation as in the previous paragraph, their anomalies are computed by the geometric invariant $\exp \pi i \xi_V$ \cite{Witten:1985xe, MR861886}. The determinant line is replaced by a Pfaffian line that can be defined using the involution.

The corresponding anomaly field theories have been constructed in Section 9 of \cite{Freed:2016rqq}.

\paragraph{Wess-Zumino terms} A Wess-Zumino term is a term in the classical action of a quantum field theory that fails to be invariant under the symmetry under consideration. Their existence can often be deduced from anomaly matching. Indeed, recall that an old argument of 't Hooft shows that the anomalies of an IR theory should coincide with the anomalies of the a UV theory from which it flows \cite{'tHooft:1979bh}. 
\footnote{Namely, one can tensor the UV theory with a compensating theory $\mathcal{C}$ (such as free fermions) to cancel the anomaly of the UV theory, ensuring that the relevant symmetry can be weakly gauged. This produces a quantum field theory, which in the IR looks like the original IR theory tensored with $\mathcal{C}$, weakly gauged. The consistency of this theory in the IR ensures that $\mathcal{C}$ cancels the anomalies of the IR theory as well, which therefore have to coincide with the anomalies of the UV theory.}
In certain cases, the anomalies of the IR theory computed from the minimal Lagrangian appear to differ from those of the UV theory. Consistency is restored by the presence of extra terms in the IR theory action, the Wess-Zumino terms, making its anomaly coincide with the UV anomaly. See \cite{Wess:1971yu, Witten:1983tw} for the original example in the non-linear sigma model describing pions.

The Green-Schwarz mechanism is a particular instance of the phenomenon described above, where the UV theory is anomaly-free and the apparent anomalies of the IR theory are canceled by Wess-Zumino terms. It will be discussed further below in the case of six-dimensional supergravity theories.

Wess-Zumino terms can be construed as follows in the formalism introduced above. Let $\mathcal{A}'$ be a $d+1$-dimensional invertible quantum field theory. Given a quantum field theory $\mathcal{F}$ with anomaly field theory $\mathcal{A}$, suppose that we manage to construct a vector $\mathcal{W}(M) \in \mathcal{A}'(M)$ for every $d$-dimensional spacetime manifold $M$, such that
\be
\label{EqFormWZTerms}
\mathcal{W}(M) = \exp 2\pi i \int_M w \;,
\ee
where $w$ is a local expression in the fields of $\mathcal{F}$. $\mathcal{W}(M) \in \mathcal{A}'(M)$ implies in particular that under the symmetry of interest, $\mathcal{W}(M)$ transforms by the character associated with the representation $\mathcal{A}'(M)$. Adding the Wess-Zumino term $w$ to the action of $\mathcal{F}$ then changes the anomaly field theory of $\mathcal{F}$ from $\mathcal{A}$ to $\mathcal{A} \otimes \mathcal{A}'$. 


\paragraph{Self-dual fields} Self-dual abelian (higher) gauge fields, also known as chiral form fields, are abelian $p$-form gauge fields on a $2p+2$-dimensional spacetime manifold, whose degree $p+1$-form field strength obey a self-duality condition. Real self-dual fields in Lorentzian signature exist only for $p = 2\ell$, $\ell = 0,1, ...$. Locally, the exterior cotangent bundle of spacetime $\bigwedge T^\ast M$ can always decomposed as the tensor product of two spin bundles $S \otimes S$. (This decomposition holds globally if and only if the manifold is spin.) In turn, the spin bundle on manifolds of dimension $4\ell+2$ decomposes as $S = S_+ \oplus S_-$, corresponding to two inequivalent representations of the Lorentz group. Chiral forms of all even degrees can be seen as sections of $S \otimes S_+ \subset S \otimes S$, and can therefore be treated formally exactly like chiral fermions valued in the chiral spin bundle $S_+$. As only the forms of degree $2\ell$ are anomalous, this implies that the geometric invariant computing the anomalies of self-dual fields is $\exp \frac{\pi i}{2} \xi_\sigma$, where $\xi_\sigma$ is the modified eta invariant of the signature Dirac operator in dimension $4\ell+3$ \cite{Witten:1985xe}. The factor $1/4$ in the exponent compared to the geometric invariant for complex chiral fermion decomposes into a factor of $1/2$ due to the reality condition satisfied by the self-dual fields in Lorentzian signature, and into a factor $1/2$ due to the fact that the twist bundle of the signature operator is the $4\ell+3$-dimensional spin bundle, which restricts to two copies of the chiral spin bundle in dimension $4\ell+2$.

The signature Dirac operator in dimension $4\ell+3$ can be expressed in terms of the Hodge star operator and of the differential as $(-1)^{\ell+p+1}(d\ast - \ast d)$ on forms of degree $2p$. This means that its kernel is given by harmonic forms and has constant dimension as the spacetime metric is varied. As a result, the eta invariant varies continuously over the space of metrics and $\exp \frac{\pi i}{2} \xi_\sigma$ is well-defined (despite the factor 1/4 mentioned above). $\exp \frac{\pi i}{2} \xi_\sigma$ can be promoted to the partition function of a "quarter Dai-Freed theory" , which again exists only thanks to the fact that the kernel of the signature Dirac operator has constant dimension. This is the anomaly field theory of self-dual fields.

There is a natural way of coupling a degree $2\ell$ self-dual gauge field to a degree $2\ell+1$ abelian gauge field \cite{Witten:1996hc, Belov:2006jd}, and such a coupling changes the anomalies of the self dual field in a non-trivial way \cite{Monnier2011a, Monniera}. The $4\ell+3$-dimensional geometric invariant gets multiplied by the exponentiated Arf invariant of a certain quadratic refinement of the linking pairing. It is easier to obtain explicit expressions for this geometric invariant when the $4\ell+3$-dimensional manifold $U$ is the boundary of a $4\ell+4$-dimensional manifold $W$. In this case, we have
\be
\label{HalfModEtaInv}
\frac{1}{4}\xi_\sigma(U) = \frac{1}{8} \left( \int_W [L(W)]_{4\ell+4} - \sigma_W \right) \;,
\ee
where $L(W)$ is the Hirzebruch L-genus, $[\bullet]_{4\ell+4}$ denotes the $4\ell+4$-form component and $\sigma_W$ is the signature of the lattice $H^{2\ell+2}_{\rm free}(W,\partial W; \mathbb{Z})$. This is the geometric invariant governing the anomaly of an uncharged self-dual field. A self-dual field with unit charge corresponds to the geometric invariant
\be
\int_W \left[ \frac{1}{8} L(W) - \frac{1}{2} G^2 \right]_{4\ell+4} \;,
\ee
where $G$ is the extension to $W$ of the $2\ell+2$-form field strength of the background gauge field coupling to the self-dual field. Crucially, $G$ has to be extended to $W$ subject to a fractional flux quantization law: the periods of $2G$ modulo 2 coincide with the periods of a certain $\mathbb{Z}_2$-valued characteristic class of $W$, the degree $2\ell+2$ Wu class of $W$. The net anomaly due to the coupling the self-dual field to the background gauge field is therefore
\be
\label{EqNetAnomCouplSDF}
\frac{1}{8} \sigma_W - \frac{1}{2} \int_W  G^2 \;.
\ee

At the level of anomaly field theories, the quarter Dai-Freed theory whose partition function is \eqref{HalfModEtaInv} gets tensored by a certain Wu Chern-Simons theory \cite{Monniere}, whose partition function on a $4\ell+3$ manifold $U$ such that $U = \partial W$ is \eqref{EqNetAnomCouplSDF}. Wu Chern-Simons theories \cite{Monnier:2016jlo} can be seen as quadratic Chern-Simons theories of degree $2\ell+1$ abelian gauge field at "half-integer level", thereby generalizing spin Chern-Simons theories to higher dimension. The definition of the anomaly field theory of charged self-dual fields is implicit in \cite{Monniere} and will be discussed in more detail elsewhere.

The above is easily generalizable to the case of self-dual fields valued in an arbitrary abelian group, instead of the $U(1)$ case considered so far. The abelian gauge group can be written $(\Lambda \otimes \mathbb{R})/\Lambda$, where $\Lambda$ is interpreted as the charge lattice of the self-dual fields. The gravitational part \eqref{HalfModEtaInv} of the anomaly simply gets multiplied by the signature of $\Lambda$, while the gauge part \eqref{EqNetAnomCouplSDF} becomes
\be
\label{EqNetAnomCouplSDFMult}
\frac{1}{8} \sigma_{\Lambda,W} - \frac{1}{2} \int_W  G^2 \;.
\ee
where $G$ is now a $2\ell+2$-form valued in $\Lambda \otimes \mathbb{R}$ and $\sigma_{\Lambda,W}$ is the signature of the lattice $H^{2\ell+2}_{\rm free}(W,\partial W; \Lambda)$. The flux quantization of $G$ is now the following \cite{Monnier:2016jlo}. Let $\Gamma^{(2)}$ be the quotient of $\Lambda/2\Lambda$ by the radical of the induced pairing. The reduction modulo 2 of the periods of $2G$ in $\Gamma^{(2)}$ should coincide with those of $\nu \otimes \gamma$, where $\nu$ is the degree $2\ell+2$ Wu class of $W$ and $\gamma$ is the unique characteristic element in $\Gamma^{(2)}$, i.e. $(x, x) = (x, \gamma)$ for all $x \in \Gamma^{(2)}$.

An interesting fact is that the Wu Chern-Simons theory encoding the anomaly of the coupling to the background gauge field is invertible if and only if $\Lambda$ is unimodular. It is indeed known that self-dual fields with non-unimodular charge lattice do not have a unique partition function \cite{Belov:2006jd}.

\paragraph{2-dimensional rational chiral conformal field theories} Chiral rational conformal field theories (RCFTs) provide further examples of anomalous field theories with non-invertible anomaly field theories.

In a nutshell, a chiral RCFT is determined by the data of a modular tensor category \cite{Moore:1988qv, 2005PNAS..102.5352H, Fuchs:2002cm, Fuchs:2004xi}, which itself can be used to construct a 3d Reshetikhin-Turaev (RT) theory \cite{Reshetikhin1991, turaev1994quantum}. The RT theory is the anomaly field theory of the rational conformal field theory. A concrete example is when the modular tensor category is the category of positive energy representations of a loop group central extension. The chiral RCFT is then a chiral WZW model whose target group and level are determined by the central extension. The anomaly field theory is a quantum Chern-Simons theory whose gauge group and level are determined similarly. See Section 3.5 of \cite{Monnierd} for more details.

This is consistent with the well-known fact that RCFTs generally do not have a single partition function, but rather a vector of " conformal blocks" that take value in the state space of the RT theory. (Indeed, this is exactly how the state space of quantum Chern-Simons theory was constructed in \cite{Witten:1988hf}.)

A common way to cancel the (non-invertible) anomalies of RCFTs is to pair positive chirality and negative chirality theories. The motivation in the literature is to find a modular invariant torus partition function. But in the framework developed above, modular invariant theories are precisely theories whose global gravitational anomalies vanish on the torus.

We should also mention that the restriction to \emph{rational} chiral conformal field theories is crucial. Treating  the non-rational case would require the anomaly field theory to have infinite dimensional state spaces, and would force us to deal with infinite dimensional 2-Hilbert spaces in dimension $1$.

\paragraph{6-dimensional (2,0) superconformal field theories} 6-dimensional (2,0) superconformal field theories (6d SCFTs, see for instance Section 6 of \cite{Moore2012}) are also examples of anomalous quantum field theories with non-invertible anomaly field theories. As they do not admit a semi-classical limit, these theories are rather hard to study and are mainly known through their string theory constructions, which nevertheless yield enough information to construct their anomaly field theories \cite{Monnier:2017klz}. Indeed, the local \cite{Witten:1996hc, Harvey:1998bx} and global \cite{Monnier:2014txa} anomalies of 6d SCFT in the A series can be computed by considering anomaly cancellation on stacks of M5-branes in M-theory. The worldvolume of a stack of $n$ M5-branes is described in a certain decoupling limit by a $A_n$ 6d SCFT and a free tensor multiplet representing the center of mass. Assuming M-theory to be consistent, and therefore non-anomalous, the anomalies of the 6d SCFT have to cancel against the anomaly inflow coming from the 11-dimensional Chern-Simons term, which can be computed explicitly. This yields the geometric invariants computing the anomalies of the 6d SCFTs in the A-series \cite{Monnier:2014txa}. Natural expressions for the geometric invariants of 6d SCFTs in the D- and E-series can be conjectured by rewriting the A-series invariant in terms of Lie algebra data. These geometric invariants can then be promoted to field theory functors \cite{Monnier:2017klz}. 

The anomaly field theory is a product of several factors, associated to fermions, self-dual fields and Hopf-Wess-Zumino terms \cite{Intriligator:2000eq}. We will not make them explicit here, see \cite{Monnier:2017klz}. Remarkably, the only non-invertible factor is a Wu Chern-Simons theory \cite{Monnier:2016jlo}, whose state space hosts the conformal blocks of the 6d SCFTs. Understanding the action of the diffeomorphisms and R-symmetry on this state space would be a step toward understanding the transformations of the conformal blocks of the 6d SCFT.

\section{Recovering the classical picture of anomalies}

\label{ClassPictAnom}

In this section, we explain how many well-known properties of anomalies and anomalous field theories are recovered in the formalism of Section \ref{AnomRelFieldTh}. As before, we consider a $d$-dimensional anomalous theory 
$\mathcal{F}: \mathcal{A} \rightarrow \bm{1}^{d+1}$ with anomaly field theory $\mathcal{A}$.

\paragraph{Local anomaly} Let us first see how to recover the local anomaly of $\mathcal{F}$. Recall that the local anomaly is completely determined by a characteristic form of degree $d+2$, the \emph{anomaly polynomial}. To recover it, we take a $d$-dimensional manifold $M$ and we consider the partition function of $\mathcal{A}$ on $M \times S^1$, with the field theory data pulled back from $M$. Then $M \times S^1$ endowed with the field theory data bounds $M \times D^2$, and for all known anomaly field theories $\mathcal{A}$, $\frac{1}{2\pi i} \ln \mathcal{A}(M \times S^1)$ modulo 1 can be expressed as the integral of a degree $d+1$ characteristic form on $M \times D^2$ up to locally constant terms. This characteristic form is the anomaly polynomial of $\mathcal{F}$.

For instance, in the case of complex chiral fermions discussed in Section \ref{Examples}, $\frac{1}{2\pi i} \ln \mathcal{A}(M \times S^1)$ is the modified eta invariant $\xi_V$, which according to the Atiyah-Patodi-Singer theorem \cite{Atiyah1973} can be written as 
\be
\xi(M \times S^1) = \int_{M \times D^2} I_V - {\rm index}(D_{V, M \times D^2}) \;,
\ee
$D_{V, M \times D^2}$ is the Dirac operator with twist $V$ on $M \times D^2$ and $I_V$ is its index density. The second term is an integer, hence irrelevant modulo 1, and we find that the anomaly polynomial of the complex chiral fermion is given by the degree $d+2$ component of $I_V$, as is well-known \cite{AlvarezGaume:1983ig}.

\paragraph{Action of symmetries on the partition function} Suppose that the anomalous field theory $\mathcal{F}: \mathcal{A} \rightarrow \bm{1}^{d+1}$ has a symmetry group $G$. This means that $\mathcal{A}$ admits $G$ as a symmetry group in the usual quantum field theory sense, that $G$ acts trivially on $\bm{1}^{d+1}$ and that $\mathcal{F}$ is equivariant:
\be
\label{EqTransPartFuncUndSym}
\mathcal{F}(\phi M^{d}) = \mathcal{A}(\phi) \circ \mathcal{F}(M^{d})
\ee 
where we used the fact that $\bm{1}^{d+1}(\phi)$ is the identity map.

Now $\mathcal{A}(M^d)$ is a representation of $G$, so $\mathcal{F}(M^d) \in \mathcal{A}(M^d)$ transforms as a vector in a representation of $G$. This is indeed the content of \eqref{EqTransPartFuncUndSym}. We just derived the fact that the partition function of an anomalous field theory is not invariant under symmetries, but only covariant.

If $\mathcal{A}$ is invertible, as in the case of chiral fermions, $\mathcal{A}(M^d)$ is 1-dimensional, and $\mathcal{F}(M^d)$ transforms by a phase. We can see $\mathcal{A}(M^d)$ as a group character, or equivalently a 1-cocycle for the group cohomology of $G$ valued in $U(1)$. The anomaly is trivial when this group character is trivial, or equivalently when the associated group cohomology class is trivial. When $G$ is a Lie group, the cocycle condition for the associated Lie algebra cocycle is known in the physics literature as the Wess-Zumino consistency conditions.

We can now understand the prescription to compute anomalous phases illustrated in Figure \ref{FigGlobAnComp}. $\mathcal{F}(M^d)$ is a vector in the 1-dimensional vector space $\mathcal{A}(M^d)$. The anomalous phase is given by the scalar product of $\mathcal{F}(M^d)$ with $\mathcal{F}(\phi M^d) = \mathcal{A}(\phi) \circ \mathcal{F}(M^d)$. But as $\mathcal{A}(M^d)$ is 1-dimensional, we can compute this scalar product as $(v, \mathcal{A}(\phi) v)$ for any unit vector $v \in \mathcal{A}(M^d)$. Such a unit vector can easily be constructed as $\mathcal{A}(N)$ where $N$ is a $d+1$-dimensional manifold such that $\partial N = M^d$. The scalar product is then computed by $\mathcal{A}$ evaluated on the twisted double manifold constructed from $N$ and $\phi$, as in the prescription of Figure \ref{FigGlobAnComp}.

In general, the partition function is vector-valued and transforms in the representation of $G$ defined by the state space of $\mathcal{A}$. An analogous cohomological interpretation of the anomaly exists: a representation can be thought of as a "non-abelian group 1-cocycle" and the triviality of the associated cohomology class is equivalent to the representation being isomorphic to the trivial one. For instance, in the case of rational chiral conformal field theories  on $T^2$ (see Section \ref{Examples}), the transformations of the conformal blocks under the diffeomorphism group are encoded into the modular $S$ and $T$ matrices. They define a representation of the diffeomorphism group constant on the connected component of the identity. 

\paragraph{Action of symmetries on the state space} We keep the same setup as in the previous paragraph, but now consider $d-1$-dimensional manifolds. $G$ also has an action on the 2-Hilbert space $\mathcal{A}(M^{d-1})$. Assuming $\mathcal{A}$ invertible, $\mathcal{A}(M^{d-1}) \simeq \mathcal{H}$ and $G$ acts on $\mathcal{H}$ through the tensor product of lines:
\be
g \cdot V = L_g \otimes V \;, \quad g \in G \;, \; V \in \mathcal{H} \;.
\ee
The group law requires isomorphisms
\be
\label{EqIsomGrpLaw}
L_{g_1g_2} \otimes L_{g_1^{-1}} \otimes L_{g_2^{-1}} \simeq \mathbb{C} \;.
\ee
After choosing non-canonical isomorphisms $L_g \simeq \mathbb{C}$, the isomorphisms \eqref{EqIsomGrpLaw} yield a $U(1)$-valued group 2-cocycle $\alpha = \{\alpha_{g_1, g_2}\}$.

$\mathcal{F}(M^{d-1}) \in \mathcal{A}(M^{d-1})$ can be (non-canonically) identified with a Hilbert space $H \in \mathcal{H}$. The action of $G$ is $g . H = L_g \otimes H$. After the identification $L_g \simeq \mathbb{C}$ we get an endomorphism $\phi_g$ of $H$ for each $g$. We also have $(g_1 g_2) . g_2^{-1} . g_1^{-1} . H = L_{g_1g_2} \otimes L_{g_1^{-1}} \otimes L_{g_2^{-1}} \otimes H$, so we deduce that
\be
\phi_{g_1 g_2} \circ \phi_{g_2^{-1}} \circ \phi_{g_1^{-1}} = \alpha_{g_1, g_2} \mathbbm{1}_H
\ee
$\phi$ is therefore a \emph{projective} representation of $G$ on $H$. We recovered the main characteristic of Hamiltonian anomalies \cite{Faddeev:1984jp, Mickelsson1985}.

There are many arbitrary choices of isomorphisms appearing in the derivation of the Hamiltonian anomaly. One can show that changes of such choices change the degree 2 cocycle $\alpha$ by a boundary, so the anomaly is characterized by the group cohomology class of $\alpha$.

The story above generalizes to non-invertible anomalies. $\mathcal{F}(M^{d-1})$ is now a vector of Hilbert spaces and the action of $G$ is through "permutation matrices of lines". To make this concrete, imagine that $\mathcal{F}(M^{d-1})$ is a vector with components $H_1, H_2 \in \mathcal{H}$. The action of $\phi \in G$ could for instance be as follows:
\be
\phi . \left ( \begin{array}{c} H_1 \\ H_2 \end{array} \right ) = \left ( \begin{array}{cc} 0 & L^{12}_\phi \\ L^{21}_\phi & 0 \end{array} \right ) \otimes \left ( \begin{array}{c} H_1 \\ H_2 \end{array} \right ) = \left ( \begin{array}{c} L^{12}_\phi \otimes H_2 \\ L^{21}_\phi \otimes H_1 \end{array} \right ) \;,
\ee
where $L^{12}_\phi$ and $L^{21}_\phi$ are hermitian lines (1-dimensional Hilbert spaces). Making trivializing choices, we can extract a non-abelian group cocycle, whose non-abelian cohomology class \cite{2006math.....11317B} characterizes the anomaly, see Section 3.4 of \cite{Monnierd} for details.

These "non-invertible Hamiltonian anomalies" have not been described in the physics literature, but should be relevant to $(2,0)$ superconformal field theories in six dimensions (see Section \ref{Examples}).

\paragraph{Geometric interpretation of anomalies} A well-known geometric interpretation of anomalies is as follows. Suppose for definiteness that we are looking at chiral fermions coupled to a gauge field. We can imagine performing the path integral for this theory in stages. First, we can integrate out the fermionic degrees of freedom, and then integrate the result over the moduli space of connections parametrizing the gauge inequivalent configuration of the gauge field.

In the presence of an anomaly of the chiral fermions, the second step is impaired by the fact that the fermionic partition function is not a function over the moduli space of connections, but rather the section of a line bundle, the anomaly line bundle. To perform the integral, we need a geometric trivialization of the anomaly line bundle, i.e. an isomorphism with the trivial bundle, which maps sections to functions. The obstruction to such a geometric trivialization is the anomaly.

This situation can be easily interpreted in the present framework. Like the anomalous chiral fermionic field theory $\mathcal{F}$, the associated anomaly field theory $\mathcal{A}$ depends on the gauge field $A$. $\mathcal{A}(M^d, A)$ is a Hermitian line (1d Hilbert space) for each $A$. When $A$ varies, $\mathcal{A}(M^d, A)$ therefore defines a Hermitian line bundle over the moduli space of gauge connections: this is the anomaly line bundle. Any path in the moduli space of gauge connection defines a cylinder $M^d \times I$ endowed with a background gauge field $A$. We can further take an "adiabatic" limit in which the Riemannian metric on the cylinder is such that it becomes infinitely long for a fixed fiber volume. The anomaly field theory evaluated on the adiabatic limit of the cylinder is a homomorphism between the Hermitian lines over the endpoints of the path. The results of \cite{Witten:1985xe, MR861886} show that this parallel transport data defines the connection on the anomaly line bundle.

\paragraph{Condition for anomaly cancellation} The present framework provides a simple condition for the cancellation of all anomalies. Indeed, we saw that a natural transformation $\mathcal{F}: \mathcal{A} \rightarrow \bm{1}^{d+1}$ is equivalent to an ordinary (i.e. anomaly free) quantum field theory when $\mathcal{A}$ is the trivial field theory functor $\bm{1}^{d+1}$. This is also true whenever $\mathcal{A}$ is naturally isomorphic to $\bm{1}^{d+1}$, as we can precompose $\mathcal{F}$ with the natural isomorphism to see it as a natural transformation from $\bm{1}^{d+1}$ to itself.

Therefore \emph{$\mathcal{F}$ is non-anomalous if and only if $\mathcal{A}$ is naturally isomorphic to $\bm{1}^{d+1}$.}

\paragraph{Setting the quantum integrand} In certain theories, anomalies cancel, but defining their $\mathbb{C}$-valued partition functions on $d$-dimensional manifolds that are not boundaries of $d+1$-dimensional manifolds require extra choices, known as the "setting of the quantum integrand" \cite{Witten:1996hc, Freed:2004yc, WittenTalkStrings2015}. More precisely, the available choices form a torsor over ${\rm Hom}({\rm Bord}_d, U(1))$, where ${\rm Bord}_d$ is the bordism group of $d$-dimensional manifolds endowed with field theory data.

This situation occurs when the anomaly field theory $\mathcal{A}$ is trivial, but not canonically so. Assuming $\mathcal{A}$ is trivial, let us think about what freedom we have in choosing an isomorphism $\mathcal{A} \simeq \bm{1}^{d+1}$. Being trivial, $\mathcal{A}$ automatically take value $1$ on every closed $d+1$-dimensional manifolds. On a $d+1$-dimensional manifold with boundary $M^{d+1,1}$, we have $\mathcal{A}(M^{d+1,1}) \in \mathcal{A}(\partial M^{d+1,1})$, while $1 = \bm{1}^{d+1}(M^{d+1,1}) \in \bm{1}^{d+1}(\partial M^{d+1,1}) = \mathbb{C}$. We see therefore that any isomorphism $\mathcal{A} \simeq \bm{1}^{d+1}$ will have to send $\mathcal{A}(M^{d+1,1})$ to $1 \in \mathbb{C}$, thereby defining a canonical trivialization of $\mathcal{A}(\partial M^{d+1,1})$. Therefore, the only freedom there is in choosing the isomorphism $\mathcal{A} \simeq \bm{1}^{d+1}$ is a trivialization of $\mathcal{A}(M^{d})$ for each $M^d$ in a basis of the $d$-dimensional cobordism group of manifolds endowed with field theory data. We precisely recovered the freedom of setting the quantum integrand of the theory $\mathcal{F}$.

\section{A concrete use case: anomaly cancellation in 6d supergravity}

\label{UseCase}

We now briefly review \cite{Monniere}, in which steps toward a systematic understanding of global anomaly cancellation in six-dimensional supergravity theories were taken (see also \cite{MonnierMooreSum2018, Moorea}).

We do not consider here six-dimensional supergravity theories whose low energy limit contains strongly coupled sectors \cite{DelZotto:2014fia}: we assume that the field content reduces to a set of free supermultiplets. 6d supergravity theories contain chiral fermions and self-dual 2-forms, and are therefore generically anomalous. A version of the Green-Schwarz mechanism can cancel the anomalies in certain cases, as explained in \cite{Sagnotti:1992qw, Sadov:1996zm}. In these classical works however, the Green-Schwarz terms are written under the assumption that the topologies of spacetime and of the gauge bundles of the 6d supergravity are trivial. A prerequisite to the study of global anomalies is therefore to understand the Green-Schwarz mechanism in topologically non-trivial setups.

\paragraph{The Green-Schwarz mechanism in 6d supergravity} The "bare" supergravity theory, i.e. before the addition of the Green-Schwarz terms, is a collection of free supermultiplets containing chiral fermions and self-dual 2-forms. The self-dual 2-forms come from the gravitational multiplet and from the tensor multiplets. The associated abelian gauge group can be written $(\Lambda \otimes \mathbb{R})/ \Lambda$ for $\Lambda$ an integral lattice, which can be interpreted as the lattice of possible charges. A necessary condition for the cancellation of local anomalies through the Green-Schwarz mechanism is that the degree 8 anomaly polynomial of the bare supergravity theory factorizes as follows:
\be
I_8 = \frac{1}{2} Y \wedge Y \;,
\ee
where $Y$ is a degree 4 $\Lambda \otimes \mathbb{R}$-valued differential form, and the wedge product above includes implicitly the lattice pairing. The vector multiplets contain gauge fields, whose associated gauge group $G$ is a compact Lie group. $G$ is of the form $G = (G_{\rm ss} \times G_{\rm ab})/\Gamma$, where $G_{\rm ss}$ is the semisimple part, $G_{\rm ab}$ the abelian part and $\Gamma$ is a discrete subgroup of the center of $G$. $Y$ can be expressed in terms of the 1st Pontryagin form $p_1$ of the spacetime, of the second Chern forms $c_2^i$ associated to the simple factors in $G_{\rm ss}$ and of the first Chern forms $c_1^I$ associated to the $U(1)$ factors in $G_{\rm ab}$ as follows:
\be
\label{EqRefEffAbGaugFFT}
Y = \frac{1}{4} a p_1 - \sum_i b_i c_2^i + \frac{1}{2} \sum_{IJ} b_{IJ} c_1^I c_1^J \,.
\ee
$a, b_i, b_{IJ} \in \Lambda \otimes \mathbb{R}$ are the \emph{anomaly coefficients} of the 6d supergravity.

Given the expressions of the geometric invariants encoding the anomalies of chiral fermions and self-dual fields, it is easy to compute the geometric invariant encoding the total anomaly of the bare 6d supergravity, which reads
\be
\label{EqAnBare6dSugra}
\frac{1}{2} \int_W Y_W \wedge Y_W - \frac{1}{8}\sigma_{\Lambda, W}
\ee
on a 7-manifold $U$ such that $U = \partial W$. $Y_W$ is here the characteristic form \eqref{EqRefEffAbGaugFFT}, computed on $W$. Comparing with \eqref{EqNetAnomCouplSDFMult}, we see that we should be able to cancel the global anomaly of the bare 6d supergravity theory by coupling the self-dual 2-form fields to an effective degree 3 abelian gauge field whose field strength is $Y$. This is indeed how the Green-Schwarz mechanism operates \cite{Sagnotti:1992qw, Sadov:1996zm}.

\paragraph{Constraints from the existence of the Green-Schwarz term} To write the Green-Schwarz term coupling the self-dual 2-forms to the effective abelian gauge field on spacetimes endowed with vector multiplet gauge bundle of arbitrary topology, we need to construct the Wu Chern-Simons theory whose partition function on a 7-manifold $U$ that bounds is given by \eqref{EqAnBare6dSugra}. Indeed, as explained in Section \ref{Examples}, the exponentiated Green-Schwarz term should be a vector in the state space of the Wu Chern-Simons theory.

The very existence of this Wu Chern-Simons theory puts rather strong constraints on the 6d supergravity theory.
\begin{enumerate}
\item The anomaly coefficients $a, b_i, \frac{1}{2} b_{II}, b_{IJ}$ must all be element of the lattice $\Lambda$, as opposed to arbitrary elements of the vector space $\Lambda \otimes \mathbb{R}$. In fact, a stronger condition depending on the global form of the gauge group $G$ holds, see \cite{Monnier:2017oqd, Monniere} for details.
\item $a$ cannot be an arbitrary element of $\Lambda$: it has to be a characteristic element, i.e. an element such that $(x,x) = (x,a)$ modulo 2 for all $x \in \Lambda$.
\end{enumerate}
Moreover, the cancellation of anomalies is possible in the present context only if the Wu Chern-Simons theory is invertible, which is the case if and only if the lattice $\Lambda$ is unimodular. We therefore recover the condition of unimodularity on $\Lambda$ that was previously derived in \cite{SeibergTaylor} using other arguments.

These constraints can be checked in 6d supergravity theories obtained as effective low energy theories from string theory solutions. As far as is known, all six-dimensional supergravity theories coming from string theory can be realized in F-theory, so it is sufficient for us to check these constraints in F-theory solutions. In F-theory, $\Lambda$ is the second homology lattice of the base of the six-dimensional elliptically fibered Calabi-Yau on which F-theory is compactified. It is therefore automatically unimodular. The anomaly coefficients have natural interpretations as divisors in the base of the fibration, so their homology classes are automatically elements of $\Lambda$. Moreover, $a$ corresponds to the canonical class of the base, and a simple use of the adjunction formula shows that it is necessarily a characteristic element of $\Lambda$ \cite{Monnier:2017oqd}. All the constraints are therefore satisfied in effective six-dimensional supergravity theories coming from string theory solutions, albeit in a rather non-trivial way.

The second constraint above allows in particular to rule out supergravity theories that so far looked perfectly consistent. Consider for instance \cite{Monnier:2017oqd} a theory with a single tensor multiplet, a charge lattice $\Lambda = \mathbb{Z}^2$ with pairing
\be
\begin{pmatrix}
0&1\\1&0
\end{pmatrix} \;,
\ee
$a =(4,1)$, no gauge symmetry and $244$ neutral hypermultiplets. This theory satisfies all the anomaly constraints that were known so far, but it turns out to be impossible to construct well-defined Green-Schwarz terms on manifolds of arbitrary topology, because $a$ is not a characteristic element of $\Lambda$. Anomalies can therefore not be canceled and the theory is inconsistent.

\paragraph{Constraints from global anomaly cancellation} Even when the Green-Schwarz terms can be constructed, global anomalies may still impose residual constraints. Indeed, in order to ensure the cancellation of all global anomalies, the anomaly field theory $\mathcal{A}$ of the bare supergravity theory has to coincide with the Wu Chern-Simons theory $\mathcal{A}_{\rm CT}$ encoding the anomaly of the Green-Schwarz term. Both theories have a partition function given by \eqref{EqAnBare6dSugra} on 7-manifolds $U$ that bound, i.e. such that there is an 8-manifold $W$ with $\partial W = U$. But depending on the gauge group (which enters the field theory data of the supergravity theory), there may be 7-manifolds that do not admit any such $W$, or more precisely that are not cobordant to zero. On such 7-manifolds, $\mathcal{A}_{\rm CT}$ may differ from $\mathcal{A}$. This means that the Green-Schwarz terms can be constructed, but that they do not cancel all the global anomalies of the 6d supergravity theory, which is therefore inconsistent.

In \cite{Monniere}, we show that for gauge groups $G$ given by arbitrary products of $U(n)$, $SU(n)$ and $Sp(n)$ factors, as well as for $G = E_8$, all the 7-manifolds are bordant to zero, ensuring that $\mathcal{A} = \mathcal{A}_{\rm CT}$ and therefore that the Green-Schwarz term cancels all anomalies. In other cases, it turns out to be very hard to check the equality of $\mathcal{A}$ and $\mathcal{A}_{\rm CT}$ on manifolds not bordant to zero. Partial results were obtained in \cite{Monniere} for supergravities with discrete gauge group, in the form of constraints on the matter representations. It was checked that these constraints are satisfied in the class of F-theory models constructed in \cite{Turner}, yielding yet another consistency check on string theory.

\section{Status of anomaly cancellation in string theory}

\label{AnomCanStringTh}

As perturbative string theory is quantum gravity on the string worldsheet, the gravitational anomalies on the string worldsheet have to vanish. The cancellation is known is this context as "modular invariance" and is well-studied. 

The aim of this section is rather to list, to the best of the author's knowledge, the string theory backgrounds in which anomaly cancellation in the low energy effective theory has been fully checked (including global anomalies), as well as backgrounds where it would be interesting to perform such checks. Suggestions for additions to this section are welcome.\footnote{As a sidenote, the absence of global anomalies in the Standard Model was proven in \cite{Freed:2006mx}.} 

We enumerate the backgrounds in which global anomaly cancellation has been fully checked.
\begin{enumerate}
\item Smooth Type I without D-branes. The local anomaly cancellation was shown in Green and Schwarz's famous paper \cite{Green:1984sg}. The proof of the cancellation of global anomalies is a rather non-trivial extension and was presented in \cite{Freed:2000ta}.
\item Smooth M-theory backgrounds without five-branes. A fermionic $\mathbb{Z}_2$ global anomaly is canceled by an anomaly in the M-theory Chern-Simons term (i.e. the Chern-Simons term of 11-dimensional supergravity plus a non-minimal gauge-gravitational term), as first shown in \cite{Witten:1996md} and studied in full detail in \cite{Freed:2004yc}.
\item Smooth M-theory, type IIA and $E_8 \times E_8$ heterotic backgrounds with five-branes. The worldsheet theory of the M5-brane has a gravitational and R-symmetry anomaly, which is canceled through anomaly inflow by the M-theory Chern-Simons term. The cancellation of local anomalies was shown in \cite{Freed:1998tg}, while the cancellation of global anomalies was shown in \cite{Monnierb}.
\item Although this is not strictly speaking a string background, let us mention as well "cohomological" type IIB supergravity, i.e. the naive version of type IIB supergravity in which the Ramond-Ramond fluxes are assumed to be classified by ordinary cohomology, rather than K-theory as in the low energy limit of the type IIB superstring. The cancellation of local anomalies was shown in the classical work \cite{AlvarezGaume:1983ig}. \cite{Witten:1985xe} found a residual global anomaly when the 10-dimensional spacetime has non-trivial homology in degree 5. This was due to the failure to take into account a contribution to the global anomaly coming from the coupling of the self-dual RR 4-form fields the B-field and RR 3-form gauge field, as shown in \cite{Monnier2011a, Monniera}. 
\end{enumerate}
Moreover, partial results have been obtained in the context of 8d F-theory compactifications \cite{Garcia-Etxebarria:2017crf}, 6d F-theory compactifications \cite{KMT2, Monniere} and 4d orientifolds \cite{GatoRivera:2005qd}. Finally, here are some backgrounds where it would be worthwile to check the cancellation of global anomalies.
\begin{enumerate}
\item The low energy limit of the type IIB superstring, i.e. type IIB supergravity with the Ramond-Ramond fluxes properly classified by (twisted) K-theory. We note that there is some tension between the K-theory flux classification and S-duality. Global anomalies may shed light on this issue.
\item More generally, Type II backgrounds including orientifolds, following the precise characterization of these backgrounds in \cite{Distler:2009ri}.
\item Type IIA D-branes. Fermions on type IIA D-branes suffer from a global $\mathbb{Z}_2$ anomaly. Ideas about how this anomaly could be canceled were offered in \cite{Freed:2000tt}.
\item 6d $N=(1,0)$ F-theory compactifications with strongly coupled sectors \cite{DelZotto:2014fia}.
\end{enumerate}

\subsection*{Acknowledgements}

This work was supported in part by the grant MODFLAT of the European Research Council, SNSF grants No. 152812, 165666, and by NCCR SwissMAP, funded by the Swiss National Science Foundation. I would like to thank the organizers of the Durham Symposium "Higher Structures in M-theory" for the opportunity to present this work.

\end{document}